\input amstex
\documentstyle{amsppt}

\mag=\magstep1
\vsize=21.6truecm
\hsize=16truecm
\NoBlackBoxes
\leftheadtext{Yunbo Zeng }
\rightheadtext{Deriving $N$-soliton solutions via constrained flows}

%\NoRunningHeads

\topmatter
 \title Deriving $N$-soliton solutions via constrained flows
\endtitle
\author
Yunbo Zeng \footnote {E-mail:yzeng\@math.tsinghua.edu.cn and
yzeng\@mail.tsinghua.edu.cn} \\
{\it Department of Mathematical  Sciences,
Tsinghua University, Beijing 100084, China}
\endauthor
\endtopmatter
%%%%%%%%%%%%%%%%%%macro%%%%%%%%%%%%%%%%%
\def\p{{\partial}}
\def\la{{\lambda}}
\def\La{{\Lambda}}

\def\Th{{\Theta}}
%%%%%%%%%%%%%%%%%%%%%%%%%%%%%%%%%%%%%%%%%%

\TagsOnRight
\ \par
\ \par
{\bf Abstract.} The soliton equations can be factorized by two commuting

$x$- and $t$-constrained flows. We propose a method to derive
$N$-soliton
solutions of soliton equations directly from the $x$- and
$t$-constrained
flows.
\par

\ \par

\newpage
\subhead {1. Introduction}\endsubhead\par
In recent years much work has been devoted to the constrained flows of
soliton equations (see, for example, [1-7]). It was shown in [1-3] that
(1+1)-dimensional soliton equation can be factorized by $x$- and
$t$-constrained flow which can be transformed into two commuting $x$-
and
$t$-finite-dimensional integrable Hamiltonian systems. The Lax
representation
for constrained flows can be deduced from the adjoint representation of
the
auxiliary linear problem for soliton equations [4]. By means of the Lax
representation, the standard method in [8-10] enables us to introduce
the
separation variables for constrained flows [11-15] and to establish
the Jacobi inversion problem [13-15]. Finally, the factorization of
soliton
equations and separability of the constrianed flows allow us to find the

Jacobi inversion problem for soliton equations [13-15]. By using the
Jacobi
inversion technique [16,17], the $N$-gap solutions in term of Riemann
theta
functions for soliton equations can be obtained, namely, the constrained

flows can be used to derive the $N$-gap solution for soliton equations.
It has been believed that the constrained flows can also been used
directly
to derive the $N$-soliton solutions for soliton equations. However this
case  remains a challenging problem.\par
It is well known that trere are several methods to derive the
$N$-soliton
solution of soliton equations, such as the inverse scattering method,
the Hirota method, the dressing method, the Darboux transformation, etc.

(see, for example, [18-20] and references therein). In present paper, we

propose a method to construct directly $N$-soliton solution from two
commuting $x$- and $t$-constrained
flows. We will illustrate the method by KdV equation. The method can be
applied to other soliton equations.\par
\ \par
\subhead {2. Constrained flows}\endsubhead \par
We first recall the constrained flows and factorization of soliton
equations
by using KdV equation. Let consider the Schr$\ddot{\text o}$dinger
spectral problem
$$-\phi_{xx}+u\phi=\la\phi. \tag 2.1$$
The KdV hierarchy associated with (2.1) can be written in
infinite-dimensional
integrable Hamiltonian system [18-20]
 $$u_{t_n}=
\p_x \frac{\delta H_{n}}{\delta u},\qquad n=1,2
,\hdots, \tag 2.2$$
where
$$\frac{\delta H_{n}}{\delta u}=L^nu,\quad L=-\p_x^2+4u-2\p_x^{-1}u_x,
\quad \p_x^{-1}\p_x=\p_x\p_x^{-1}=1.\tag 2.3$$\par
The well known KdV equation reads
$$u_t-6uu_x+u_{xxx}=0.\tag 2.4$$
For KdV equation (2.4), the time evolution equation of $\phi$ is given
by
$$\phi_t=4\la\phi_x+2u\phi_x-u_x\phi. \tag 2.5$$
The compatibility condition of (2.1) and (2.5) gives rise to (2.4).\par
It is known that
$$\frac {\delta\la}{\delta u}
= \phi^2.\tag 2.6$$\par
The constrained flows of the KdV hierarchy consists
 of the equations obtained from
 the spectral problem  (2.1) for $N$
distinct real numbers $\lambda_j$ and the restriction of the variational

derivatives for the conserved
quantities $H_{k_0}$ (for any fixed $k_0$) and $\lambda_{j}$  [2-4]
$$-\phi_{j,xx}+u\phi_{j}=\la_j\phi_{j},\qquad
j=1,...,N,\tag 2.7a$$
$$\frac {\delta H_{k_0}}{\delta u}-
\sum_{j=1}^{N}\alpha_j\frac {\delta \lambda_{j}}{\delta u}
=0.\tag 2.7b$$
The system (2.7) is invariant under all the KdV flows (2.2).\par
For $k_0=0$, in order to obtain $N$-soliton solution, we take
$$\la_j<0,\qquad
 \zeta_j=\sqrt {-\la_j},\qquad \alpha_j=4\zeta_j, \qquad j=1,...,N,$$
 one gets from (2.7b)
$$u=4\sum_{j=1}^N\zeta_j\phi_j^2=4\Phi^T\Th\Phi,\tag 2.8$$
where
$$\Phi=(\phi_{1},\cdots,\phi_{N})^{T},\quad
\Th=diag(\zeta_{1},\cdots,\zeta_{N}),\quad  \Lambda=diag
(\lambda_1,\cdots,\lambda_N).$$
 By substituting (2.8),  (2.7a) becomes
$$-\phi_{j,xx}+4\sum_{i=1}^N\zeta_i\phi_i^2\phi_j=\la_j\phi_{j},\qquad
j=1,...,N,$$
or equivalently
$$ \Phi_{xx}=-\La\Phi+4\Phi\Phi^T\Th\Phi.\tag 2.9$$\par
After inserting (2.8), (2.5) reads
$$ \Phi_{t}=4\La\Phi_x+8\Phi_x\Phi^T\Th\Phi
-8\Phi\Phi^T\Th\Phi_x.\tag 2.10$$
The compatibility of (2.7), (2.10) and (2.4) ensures that  if $\Phi$
satisfies two compatible systems (2.9) and (2.10), simultaneously, then
$u$
given by (2.8) is a solution of KdV equation (2.4), namely, the KdV
equation
(2.4) is factorized by the $x$-constrained flow (2.9) and
$t$-constrained
flow (2.10).\par
The Lax representation for the constrained flows (2.9) and (2.10), which

can be deduced from the adjoint representation of the spectral problem
(2.1)
  by using the method in [3,4], is given by
  $$Q_x=[\widetilde U, Q],$$
where $\widetilde U$ and the Lax matrix $Q$ are of the form
$$\widetilde U=\left( \matrix
0&1\\-\la+4\Phi^T\Th\Phi&0\endmatrix\right),
\qquad M=\left( \matrix
A(\la)&B(\la)\\C(\la)&-A(\la)\endmatrix\right),$$
$$A(\la)=-2\sum_{j=1}^{N}\frac{\zeta_j\phi_{j}\phi_{j,x}}
{\la-\la_{j}},\quad
B(\la)=1+2\sum_{j=1}^{N}\frac{\zeta_j\phi_{j}^2}{\la-\la_{j}},$$
$$C(\la)=-\la+2\Phi^T\Th\Phi-2\sum_{j=1}^{N}\frac{\zeta_j\phi_{j,x}^2}
{\la-\la_{j}}. $$
Then  $\frac 12Tr M^2(\la)=A^2(\la)+B(\la)C(\la)$, which
is a generating function of integrals of motion for the system (2.9) and

(2.10), gives rise to
$$A^2(\la)+B(\la)C(\la)=-\la-2
\sum_{j=1}^{N}\frac{F_{j}}{\la-\la_{j}}, $$
where $F_j, j=1,...,N,$ are $N$ independent integrals of motion for the
systems (2.9) and (2.10)
$$F_j=\phi_{j,x}^2+(\la_j-2\sum_{i=1}^N\zeta_i\phi_i^2)\phi_{j}^2
+2\sum_{k\neq j}\frac{\zeta_k(\phi_{j,x}\phi_{k}
-\phi_{j}\phi_{k,x})^2}{\la_j-\la_{k}}, \qquad j=1,...,N.\tag 2.11$$\par

\ \par
\subhead{3. Deriving $N$-soliton solution}\endsubhead\par
In order to constructing $N$-soliton solution, we have to set $F_j=0$.
It follows from (2.9) that
$$\frac{\phi_{j,x}\phi_{k}-\phi_{j}\phi_{k,x}}
{\la_j-\la_{k}}=-\p_x^{-1}(\phi_{j}\phi_{k}).\tag 3.1$$
Then one gets
$$F_j=\phi_{j,x}^2+(\la_j-2\sum_{i=1}^N\zeta_i\phi_i^2)\phi_{j}^2
-2\sum_{k=1}^N\zeta_k(\phi_{j,x}\phi_{k}
-\phi_{j}\phi_{k,x})\p_x^{-1}(\phi_{j}\phi_{k})=0
, \qquad j=1,...,N.\tag 3.2$$
The integrals of motion $F_j$ can be used to reduce the order of system
(2.9).
By multiplying (2.9) by $\phi_j$ and adding it to (3.2), one obtains
$$-\phi_j[\phi_{j,x}-2\sum_{k=1}^N\zeta_k\phi_{k}
\p_x^{-1}(\phi_{j}\phi_{k})]_x
+\phi_{j,x}[\phi_{j,x}-2\sum_{k=1}^N\zeta_k\phi_{k}
\p_x^{-1}(\phi_{j}\phi_{k})]=0, \quad j=1,...,N,$$
which results to
$$\phi_{j,x}-2\sum_{k=1}^N\zeta_k\phi_{k}
\p_x^{-1}(\phi_{j}\phi_{k})=-\gamma_j\phi_j,
\qquad \gamma_j=\gamma_j(t),
\quad j=1,...,N,$$
or equivalently
$$\Phi_{x}=-\Gamma \Phi
+2\p_x^{-1}(\Phi\Phi^T)\Th\Phi, \tag 3.3$$
where $\Gamma=diag(\gamma_1,...,\gamma_N).$ Set
$$R=2\p_x^{-1}(\Phi\Phi^T)\Th, \tag 3.4$$
Eq. (3.3) can be rewritten as
$$\Phi_{x}=-\Gamma \Phi+R\Phi. \tag 3.5$$
Notice that
$$2\Phi\Phi^T=R_x\Th^{-1},\qquad
\Th R=R^T\Th, \tag 3.6$$
it follows from (3.4) and (3.5)that
$$R_x=2\p_x^{-1}(\Phi_x\Phi^T+\Phi\Phi_x^T)\Th$$
$$=2\p_x^{-1}(-\Gamma R_x+RR_x-R_x\Gamma+R_xR)
=-\Gamma R-R\Gamma+R^2.\tag 3.7$$\par
We now show that
$$\gamma_j^2=-\la_j,\qquad \text {or} \qquad \Gamma^2
=-\La. \tag 3.8$$
In fact, it is found from (3.5), (3.6) and (3.7) that
$$\Phi_{xx}=-\Gamma \Phi_x+R\Phi_x+R_x\Phi
=\Gamma^2\Phi+(-\Gamma R-R\Gamma+R^2)\Phi+R_x\Phi$$
$$=\Gamma^2\Phi+2R_x\Phi=\Gamma^2\Phi+4\Phi\Phi^T\Th\Phi,$$
which together with (2.9) leads to (3.8).
Therefore, we can take $\Gamma=\Th$, (3.5) and (3.7) can be rewritten as

$$\Phi_{x}=-\Th \Phi+R\Phi, \tag 3.9$$
and
$$R_x=-\Th R-R\Th+R^2, \tag 3.10$$
$$2\Phi\Phi^T=R_x\Th^{-1}
=-\Th R\Th^{-1}-R+R^2\Th^{-1}. \tag 3.11$$\par
To solve (3.9), we first consider the linear system
$$\Psi_x=-\Th\Psi. \tag 3.12$$
It is easy to see that
$$\Psi=(c_1(t)e^{-\zeta_1x},...,
c_N(t)e^{-\zeta_Nx})^T.\tag 3.13$$
Take  the solution of (3.9) to be of the form
$$\Phi=\Psi-M\Psi, $$
then $M$ has to satisfy
 $$M_x=-\Th M+M\Th-R+RM. \tag 3.14$$
 Comparing (3.14) with (3.10), one finds
$$M=\frac 12 R\Th^{-1}=\p_x^{-1}(\Phi\Phi^T). \tag 3.15$$
So we have
$$\Phi=(I-M)\Psi=[I-\p_x^{-1}(\Phi\Phi^T)]\Psi, \tag 3.16$$
which leads to
$$\Psi=\sum_{n=0}^{\infty}M^n\Phi. \tag 3.17$$\par
By using (3.15) and (3.17), it is found from that
$$\p_x^{-1}(\Psi\Psi^T)=\p_x^{-1}
\sum_{n=0}^{\infty}\sum_{l=0}^{n}M^l\Phi\Phi^T M^{n-l}$$
$$=\p_x^{-1}\sum_{n=0}^{\infty}\sum_{l=0}^{n}M^lM_xM^{n-l}
=\sum_{n=1}^{\infty}M^n. \tag 3.18$$
Set
$$V=(v_{ij})=\p_x^{-1}(\Psi\Psi^T), \qquad
v_{ij}=- \frac {c_i(t)c_j(t)}{\zeta_i+\zeta_j}
e^{-(\zeta_i+ \zeta_j)x}.\tag 3.19$$
One obtain
$$(I+V)\Phi=\Psi,\qquad\text {or}
\qquad \Phi=(I-M)\Psi=(I+V)^{-1}\Psi. \tag 3.20$$\par
By inserting (3.9) and (3.11), (2.10) becomes
$$\Phi_t=[4\Th^{3}-4\Th^2R+8(-\Th+R)\Phi\Phi^T\Th
-8\Phi\Phi^T\Th(-\Th+ R)]\Phi$$
$$=4\Th^3\Phi-4R\Th^{2}\Phi. \tag 3.21$$
Let $\Psi$ satisfy the linear system
$$\Psi_t=4\Th^3\Psi, \tag 3.22$$
then
$$\Psi=(c_1(t)e^{-\zeta_1x},..., c_N(t)e^{-\zeta_Nx})^T,
\qquad c_i(t)=\beta_je^{4\zeta_j^3t},
\quad j=1,...,N.\tag 3.23$$\par
We now show that $\Phi$ determined by (3.20) and (3.23) satisfy (3.21).
In fact, we have
$$\Phi_t=-(I+V)^{-1}V_t(I+V)^{-1}\Psi+(I+V)^{-1}\Psi_t$$
$$=4\Th^3\Phi-4M\Th^3\Phi-4(I-M)V\Th^{3}\Phi $$
$$=4\Th^3\Phi-8M\Th^3\Phi=4\Th^3\Phi-4R\Th^2\Phi. $$
Therefore $\Phi$ given by (3.20) and (3.23) satisfies (2.9) and (2.10)
simultaneously
and $u=4\Phi^T\Th\Phi$ is the solution of KdV equation (2.4). It is easy

to show that this solution is just the $N$-soliton solution. Notice that

$$2\p_x (\Psi^T\Phi)= -2\Psi^T\Th\Phi+2\Psi^T(-\Th+R)\Phi$$
$$=-4\Phi^T(I+V)(I-M)\Th\Phi=-4\Phi^T\Th\Phi,$$
namely
$$u=-2\p_x\sum_{i=1}^N c_i(t)e^{-\zeta_ix}\phi_i.\tag 3.24$$
Formulas (3.20), (3.23) and (3.24) are just that obtained from
the Gelfand-Levintan-Marchenko equation for determining the $N$-soliton
solution for KdV equation [17,18,19] and finally results to the
well-known
expression for $N$-soliton solution of KdV equation (2.4)
$$u=-2\p_x^2\text{ln}(\text {det}(I+V)).$$\par
\ \par
\subhead {4. Conclusion}\endsubhead\par
The factorization of the KdV equation into two compatible $x$- and
$t$-constrained
flows enables us to derive directly the $N$-soliton solution via the
$x$- and $t$-constrained flows. The method presented in this paper
can be applied to other soliton equations for directly obtaining
$N$-soliton solutions
from constrained flows.\par

\ \par
\subhead {Acknowledgments}\endsubhead\par
This work was supported by the Chinese Basic Research Project
"Nonlinear Sciences".

\ \par

\subhead {References}\endsubhead \par
\item  {1.} Cao Cewen, 1991, Acta math. Sinica, New Series 17, 216. \par

\item {2.} Zeng Yunbo, 1991, Phys. Lett. A 160, 541.\par
\item {3.} Zeng Yunbo, 1994, Physica D 73, 171.\par
\item {4.} Zeng Yunbo and Li Yishen, 1993, J. Phys.  A 26, L273.\par
\item{5.} Antonowicz M and Rauch-Wojciechowski S, 1992, Phys. Lett. A
171, 303.\par
\item{6.}Ragnisco O and Rauch-Wojciechowski S, 1992, Inverse Problems 8,
245.\par
\item {7.} Ma W X and Strampp W, 1994, Phys. Lett. A 185, 277.\par
\item  {8.}Sklyanin E K, 1989, J. Soviet. Math. 47, 2473.\par
\item  {9.}Kuznetsov V B, 1992, J. Math. Phys. 33, 3240.\par
\item  {10.}Sklyanin E K, 1995, Theor. Phys. Suppl. 118, 35.\par
\item  {11.}Eilbeck J C, Enol'skii V Z, Kuznetsov V B and Tsiganov A V,
1994, J. Phys. A 27, 567.\par
\item  {12.}Kulish P P, Rauch-Wojciechowski S and Tsiganov A V,1996,
 J. Math. Phys. 37, 3463.\par
\item  {13.}Zeng Yunbo, 1997, J. Math. Phys. 38, 321.\par
\item{14.}Zeng Yunbo, 1997, J. Phys. A 30, 3719.\par
\item {15.}Zeng Yunbo and Ma W X, 1999, Physica A 274(3-4), 1.\par
\item{16.}Dubrovin B A, 1981, Russian Math. Survey 36, 11.\par
\item{17.} Its A and Matveev V, 1975, Theor. Mat. Fiz. 23, 51.\par
\item{18.} Ablowitz M and Segur H, 1981, Solitons and the inverse
scattering transform (Philadelphia, PA:SIAM).\par
\item{19.}Newell A C, 1985, Soliton in Mathematics and Physics
(Philadelphia, PA:SIAM).\par
\item{20.}Faddeev L D and Takjtajan L A, 1987, Hamiltonian methods in
the
theory of solitons (Berlin: Springer).\par

\bye
\bye

\bye
\bye